# Catastrophe machines a few nanometers in size

*Vladik A. Avetisov, Anastasia A. Markina, Alexander F. Valov*

N. N. Semenov Institute of Chemical Physics of the Russian Academy of Sciences, 119991 Moscow, Russia

ABSTRACT. Using molecular dynamic simulations of short oligomeric fragments of thermosensitive polymers exposed to power loads, we established three effects characteristic of classical catastrophe machines such as the Euler arch. These effects include the threshold effect (smooth responses of the oligomer to external forces below the threshold load), the bifurcation effect (the emergence of a new conformational state above the critical loads), and the hysteresis effect (different values of the critical loads when moving forth and back in the parametric force space). A nanoscale Euler arch made from short oligomers demonstrates low-frequency, mechanic-like vibrations near the bifurcation points, which we associate with an effect similar to thermal activated bistability. All of these effects may be attractive for designing molecular machines and nanodevices with mechanic-like functioning.

Catastrophe machines are classical mechanical devices in which force loads applied to elastic constructions can cause abrupt changes in the constructions' shapes. A commonplace metal ruler is a simple example of a catastrophe machine known as the Euler arch. Indeed, if we slightly compress the ruler in the longitudinal direction, it will return to its straight form. However, as soon as the compressive force exceeds a threshold, the ruler will bend and will take on an arcuate shape. Abrupt changes in states, referred to as "catastrophes" in the theory of nonlinear systems, are manifested in many physical systems. The Euler arch and Zeeman's famous installation [1] are classic catastrophe machines described in all textbooks on the theory of catastrophes [2,3].

How small can a catastrophe machine be? This issue is of interest in at least two respects. On the one hand, it obviously has to do with nanomechanics, specifically, with the smallest scales on which molecular compositions still can work as elastic mechanical constructions [4,5]. On the other hand, it clearly relates to molecular structures with properties inherent to elastic mechanical elements for using such structures for the design and construction of molecular machines. Thus far, the construction of molecular machines has mainly relied on outstanding achievements in the chemical synthesis of "molecular artifacts" (see, for example, [6-8]). If catastrophe machines a few nanometers in size can be realized in practice, they would represent an attractive alternative for designing molecular devices with mechanic-like functioning.

We have attempted to shed light on the question posed above using computer simulations. To this end, we studied the response of sort oligomers to external force loads. The results turned out to be more impressive than we had expected. We found that certain oligomers of thermosensitive polymers, themselves only a few nanometers in size, responded to the force loads exactly as an Euler arch, i.e., they are catastrophe machines by themselves.

Because the Euler arch is a mechanical prototype of our target molecular structure, we will first introduce the arch construction and its response to the force loads. In its simplest incarnation, an Euler arch can be assembled from two rigid rods joined by a hinge with an elastic spring (Figure 1a). The spring prevents the arch from bending; the arch maintains a straightened shape in the absence of external loads. To demonstrate abrupt changes in the shape of the arch, a longitudinal load, $F$, and a lateral load, $G$, are applied to the arch as shown in Figure 1a. Then, the catastrophes can be easily forecast from elementary consideration of the stationary states given by the extrema of the potential energy of the system (see, for example, textbooks [2,3]).

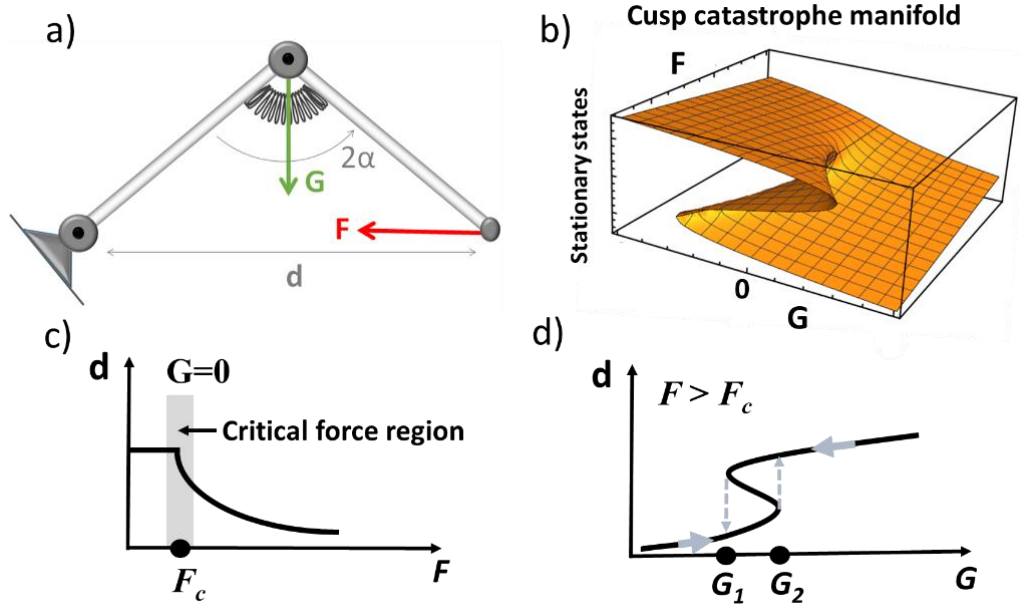

**Figure 1.** A sketch of catastrophes of the Euler arch. a) The Euler arch subject to longitudinal, $F$, and lateral, $G$, loads. b) A cusp-catastrophe manifold of the arch stationary states over the parametric space $(F,G)$. c) The bifurcation effect: $d_e$ is the distance between the arch ends; $F_c$ is the bifurcation point. d) The hysteresis effect: $G_1$ and $G_2$ are the critical points with the catastrophes when the lateral load decreases and increases, respectively.

Indeed, assuming that each rod is one unit in length, the potential energy of the arch, i.e., the sum of the elastic energy of the spring and the mechanical work of external forces applied to the arch, can be written in the following dimensionless form:

$$U(\alpha) = \frac{1}{2}\gamma(2\alpha)^2 + 2F\cos\alpha + G\sin\alpha , \qquad (1)$$

where $\gamma$ is an elastic modulus of the spring and $2\alpha$ is an angle between the rods at the spring location. Expended in a Taylor series to the 4$^{\text{th}}$ order in $\alpha$, the potential energy is

$$U(x) = x^4 - (F - 2\gamma)x^2 + Gx , \qquad (2)$$

where $x = \alpha - \frac{G}{2(2\gamma + F)}$. It should be pointed out that, in this consideration, the Euler arch is a nonlinear system with an order parameter $\alpha$ (or $x$) and control parameters $F$ and $G$. Minima in the potential energy given by the real roots of an algebraic equation

$$-\frac{dU}{dx} = -4x^3 + 2(F - 2\gamma)x - G = 0 \qquad (3)$$

describe the stationary states of the arch subject to the force loads. Depending on the values of $F$ and $G$, Equation (3) has one or three real roots. Accordingly, the system potential may have one or three extrema. A manifold $M$ of the stationary states over two-dimensional parametric space $(F, G)$ is known as the cusp catastrophe manifold and is shown in Figure 1b. As long as the longitudinal load $F$ remains less than the threshold value $F_c = 2\gamma$, Equation (3) has one real root and the surface $M$ has no folds. Accordingly, the potential energy has one minimum, and the arch smoothly responds to the loads. However, above the threshold value ($F > F_c$), the surface $M$ has a pleat over a part of the parametric space. Therefore, in this region the potential has two local minima that are both related to the bended state; the maximum between the two minima relates to the straightened state. Above the threshold load, the straightened state becomes unstable and the arch bends (Figure 1c). With the addition of lateral loads, the arch experiences an abrupt transition from one side of the pleat to another as soon as the lateral force $G$ shifts the system at an edge of the pleat (Figure 1d). It is important that each of the two sides of the pleated fold possess their own edge such that the way back to the former state is not the same way as the forward one. As a result, there is hysteresis in the transitions when moving forth and back in parametric space. The last feature makes it possible to drive the catastrophe machine in a cyclical way in parametric space to perform useful work.

In sum, three basic effects are characteristic of a classic Euler arch. i) There exists a *threshold effect*, such that no abrupt changes in arch states occur below the threshold longitudinal load. ii) There exists the *bifurcation effect*, namely, above the longitudinal threshold, the straightened state becomes unstable and the catastrophes with abrupt transitions to the bended state occur when the lateral load exceeds some critical value. iii) There exists the *hysteresis effect*, i.e., the critical loads with the catastrophes depend on the paths along which the catastrophe machine is driven in parameter space.

These three classic effects were established for molecular structures a few nanometers in size designed from short oligomers of thermosensitive polymers. Besides the classic catastrophes, the designed nanoscale Euler arch demonstrates a specific effect, which we term *thermal activated bistability*. The effect reveals itself as mechanic-like, low-frequency vibrations of the oligomer near the critical loads. This effect appears to be a size effect—in such small, nonlinear systems soft dynamical modes may be activated near the critical points by thermal fluctuations [3,9,10].

The target structures were designed following some preliminary ideas. First, we wanted the structure to be similar to the Euler arch with two rigid elongated fragments connected by a bending region. A physically attractive image is an oligomer approximately two Kuhn segments in length. We assumed that the Kuhn segment was not significantly less than 1 nanometer in length to ensure that the elongated fragments were largely immune to thermal fluctuations. We also stipulated that the structure had to possess the features of a nonlinear system with critical behavior in the sense of transitions between two well-defined conformational states. Based on these criteria, we settled on water-soluble thermosensitive polymers [11,12] as promising sources for the target oligomers.

In this letter, we announce the results related to an *N*-isopropylmethylacrylamid oligomer with $N = 30$ units in a syndiotactic configuration (oligo-30s-NIPMAm) subject to force loads. The length and the tacticity of NIPMAm-oligomer were specified by computer simulations. It should be noted that oligo-30s-NIPMAm is not a unique target sample. We found different oligomers and oligomeric compositions of thermosensitive polymers with properties of the nanoscale Euler arch. All of these structures will be reviewed in a separate paper.

The computer experiments were performed using molecular dynamic simulations (GROMACS 5.1.1 software [13-15]) of a canonical (NVT) ensemble (box size: 5.0x3.0x3.0 nm) with a time step of 1 fs. The dynamics of the target structure and the environmental water were modeled in a fully atomistic representation using the OPLS-AA force field [16,17] with the parameters presented in Figure S1 in the Supporting Information.

With the oligo-30s-NIPAm in hand, we first started with simulations of molecular dynamics at different temperatures to make sure that the target structure indeed behaved as two rigid fragments with a bending area between them. According to available experimental data, the low critical solution temperature (LCST) of the oligo-30s-NIPAm was specified to be close to 305 K [18]. Figure 2a shows the typical conformations of oligo-30s-NIPAm equilibrated sufficiently far from the LCST at 290 K and 320 K. The equilibration was controlled by the distance, $R_e$, between the oligomer ends (Figure 2b) and the gyration radius (not shown). The oligo-30s-NIPAm indeed behaves like two rigid fragments joined by a bendable area, and it has two well-distinguished conformational states, straightened and bended, at 290 K and 320 K.

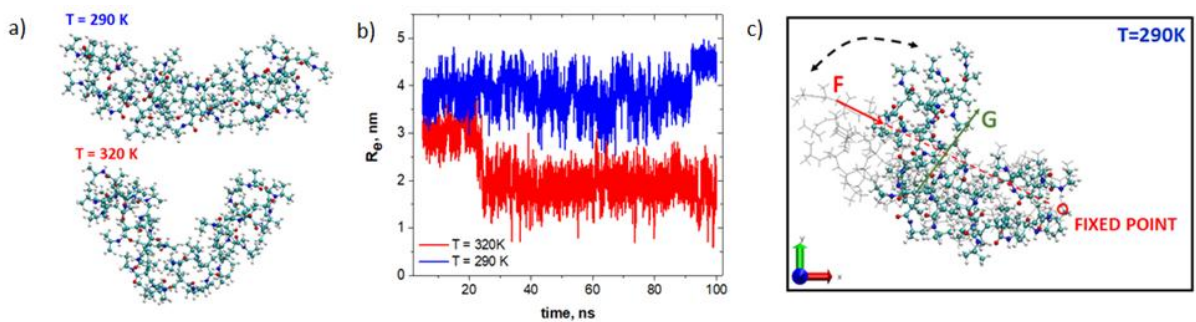

**Figure 2.** a) Typical shapes of the oligo-30s-NIPMAm after 100 ns of equilibration at temperatures T=290 K and T=320 K (water molecules not shown); b) Equilibration of the oligo-30s-NIPMAm at 290 and 320 K, respectively; the straightened and bended shapes shown in panel a) meet the trajectories shown in panel b). c) A snapshot of the oligo-30s-NIPMAm subject to a longitudinal load $F$ and a lateral load $G$.

Using the oligo-30s-NIPAm equilibrated in the straightened state (at 290 K), we applied a longitudinal force, $F$, and a lateral force, $G$, (Figure 2c) in the same manner as with the classic Euler arch (Figure 1a). The responses of the oligomer to the force loads were controlled by the end-to-end distance, $R_e$. In fact, one can take the oligomer equilibrated in the bended state (at 320 K) and apply the forces in the directions opposite to those shown in Figure 2c. We also checked this set up, and the results were qualitatively the same.

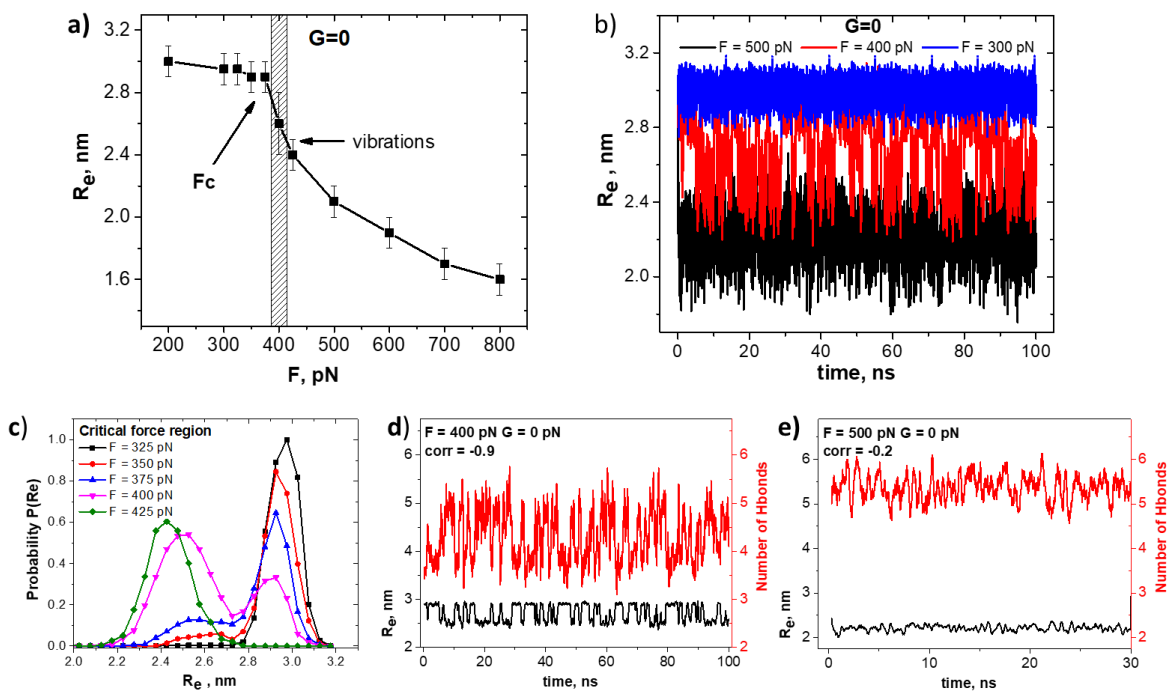

**Figure 3.** The response of the oligo-30s-NIPMAm to a longitudinal force $F$ ($G=0$). a) Stationary states (the end-to-end distance $R_e$) versus longitudinal force $F$. b) Time dependence of the distance $R_e(t)$ in equilibrated states far from the bifurcation point ($F=300$ pN and $F=500$ pN; black and blue curves, respectively) and near the bifurcation point ($F=400$ pN; red curve). c) Transformation of the probability distribution $P(R_e,F)$ of distances $R_e$ specified over the time trajectory $R_e(t)$ when $F$ passes through the bifurcation point. d) We noted a strong correlation between nanomechanical vibration near the bifurcation point ($F=400$ pN; black curve; left axis) and variations in the hydrogen bonds located in the bending area of the oligomer (red curve; right axis): the correlation coefficient was equal to $-0.9$. e) Weak correlation between the oligomer dynamics and variations in the hydrogen bonds far above the bifurcation point ($F=500$ pN); the correlation coefficient was equal to $-0.2$.

Figure 3 shows how the oligomer responds to a longitudinal force $F$ ($G$ is set to 0). Similar to the threshold load for the classical Euler arch, one can clearly see the same *threshold effect* for the oligo-30s-NIPMAm (Figure 3a). When $F$ is sufficiently less than a threshold value ($F_c \approx 400$ pN), the straightened state remains stable. Again, similar to the bifurcation of the classical Euler arch, we observe the same *bifurcation effect* for the oligo-30s-NIPMAm. As soon as the longitudinal force exceeds the threshold, the straightened state loses stability and the structure bends. Note that susceptibility to the force loads, $dR_e(F)/dF$ exhibits a discontinuity at the bifurcation point; the order parameter $R_e(F)$ changes continuously with $F$. Therefore, a new state emerges softly at the bifurcation point, which is also characteristic of the cusp-catastrophe model [3].

At the same time, near the threshold load $F_c$, the oligomer behaves in a special way. Figure 3b shows the $R_e(t)$ trajectories for loads $F$ significantly below the bifurcation point, in the vicinity of this point, and significantly above it. Sufficiently far from the bifurcation point, the oligomer standardly responds on the longitudinal load: it remains straightened below the threshold load and it bends above the threshold load. However, in the vicinity of the threshold load, the oligomer has no well-defined state: it spontaneously vibrates at a low frequency between the straightened and the bended states.

In order to specify the vibration effect more accurately, we plotted the probability distribution $P(R_e)$ of distances $R_e$ over the time trajectories $R_e(t)$ and studied how the distribution was transformed when $F$ passed through the bifurcation point (Figure 3c). We note that the inverse of the probability distribution, $[P(R_e)]^{-1}$, yields a shape of the system energy, $U(R_e)$, whose minima and maxima conversely correspond to the maxima and minima of $P(R_e)$. One can see that far from the bifurcation point from both sides the system dynamics are robust and are characterized by single-pick distributions $P(R_e)$ and, consequently, by single-well potentials $U(R_e)$ related to the straightened and bended states, respectively. However, close to the bifurcation point, the distributions $P(R_e, F \approx F_c)$ have a double-peak form, and the dynamics are controlled by a double-well potential with two minima separated by an activation barrier. In other words, near the bifurcation point, the oligomer exhibits features typical of two-state bistable systems. According to our computer experiments, the durations of the vibration pulses typically ranged from 5–10 ns (the vibration frequency varied roughly from 100–200 MHz). Therefore, the activation barrier, $\Delta H$, between two states was estimated to be roughly 10–15 $kT$ according to the familiar relation $\nu \sim 10^{13} \exp(-\Delta H/kT)$.

This estimate assumes that the nanomechanical vibrations are activated thermally (e.g., due to switching of hydrogen bonds between the oliogo-30s-NIPMAm and environmental water). To verify this idea, we checked the correlation between the nanomechanical vibrations and variations in the number of hydrogen bonds between the oligomer chain and environmental water. We found a pronounced correlation between the time series only near the bifurcation point and only for hydrogen bonds located in the bending area of the oligomer (HBBA). Figure 3d shows that the low-frequency vibrations and variation in the number of HBBA near the bifurcation point are synchronized in an antiphased manner with a correlation coefficient of $-0.9$: when the oligomer is unbent, several hydrogen bonds in the bending area are switched from the surrounding water to the oligomer, and vice versa. For comparison, the time series of HBBA and $R_e(t)$ far from the bifurcation point is shown in Figure 3e. In this case, there are no vibrations and the time series correlate poorly with one another.

Next, we turned to experiments with longitudinal and lateral forces (Figure 4). First, we made sure that the target structure subject to weak longitudinal loads $F$ smoothly responded to a lateral load $G$ (Figure 4a).

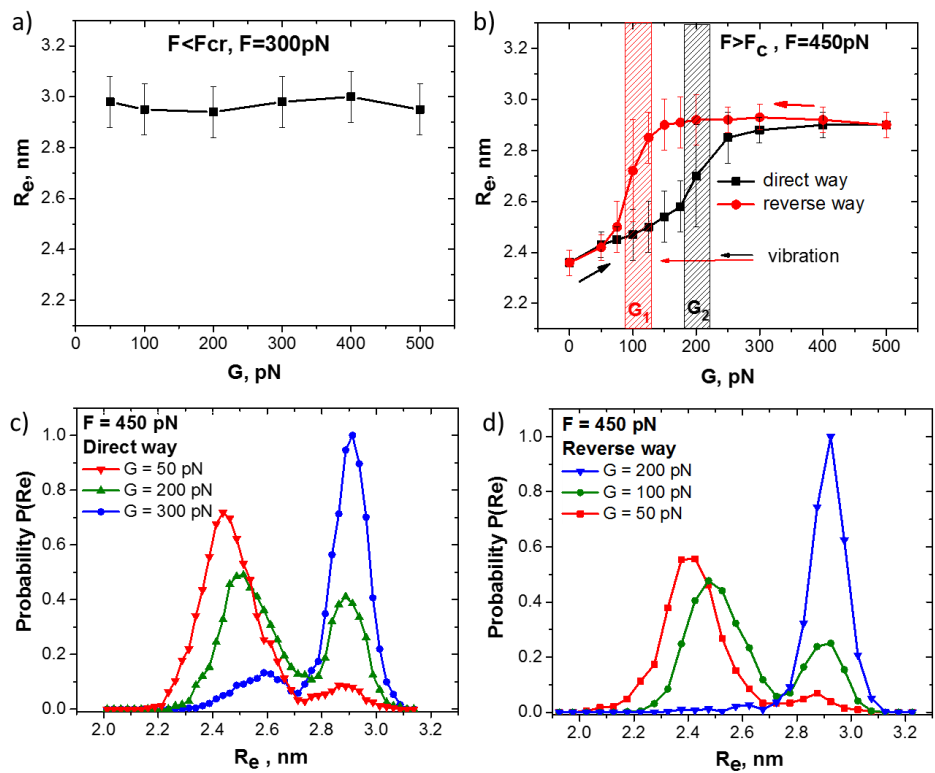

**Figure 4**. The responses of oligo-30s-NIPMAm to lateral loads $G$ below and above the threshold longitudinal force $F_c \approx 400$ pN. a) End-to-end distance $R_e$ versus lateral force $G$ below the threshold $F_c$. b) End-to-end distance $R_e$ versus lateral force $G$ above the threshold $F_c$. The black curve (marked "direct way") corresponds to an increasing lateral load applied to the oligomer initially bent due to the action of the longitudinal load. The red curve (marked "reverse way") corresponds to a decreasing lateral load applied to the oligomer initially bent due to the action of both the longitudinal and lateral loads. The critical behavior of the oligomer and the hysteresis effect can be both clearly seen. c) Variation in the probability distribution $P(R_e, G)$ under increasing $G$ along the direct way. Critical behavior with the vibration effect appears in the vicinity of the value $G_2 = 200$ pN. d) Variation of the probability distribution $P(R_e, G)$ under decreasing $G$ along the reverse way. Critical behavior with the vibration effect appears in the vicinity of the value $G_2 = 100$ pN.

However, above a threshold $F_c \approx 400$ pN, the oligomer experiences sharp changes in its states when the lateral load $G$ passes through some critical regions (Figure 4b). As soon as the increasing lateral load exceeds $G_2 \approx 200$ pN, the bended state becomes unstable and the oligomer sharply transits to the straightened state; the back transition to the bended state via a decrease in $G$ occurs at a value $G_1 \approx 100$ pN. Therefore, similar to the hysteresis effect for the classic Euler arch, one can clearly observe the same *hysteresis effect* for the oligo-30s-NIPMAm.

Furthermore, similar to the bistability near the $F_c$ threshold, the transitions between the straightened and bended states near the critical lateral loads $G_1$ and $G_2$ also occur via bistability (Figures 4c,d). Far from the critical loads from both sides, the system dynamics are robust and are characterized by single-pick distributions $P(R_e)$ and, consequently, by single-well potentials $U(R_e)$ related to the straightened and bended states, respectively. Close to the critical points, the distributions $P(R_e, G \approx G_1)$ and $P(R_e, G \approx G_2)$ have a double-peak form, and the dynamics are controlled by double-well potentials with two minima separated by an activation barrier. Figure 5 shows the vibration effect in the vicinity of the critical lateral loads and far from the critical lateral loads.

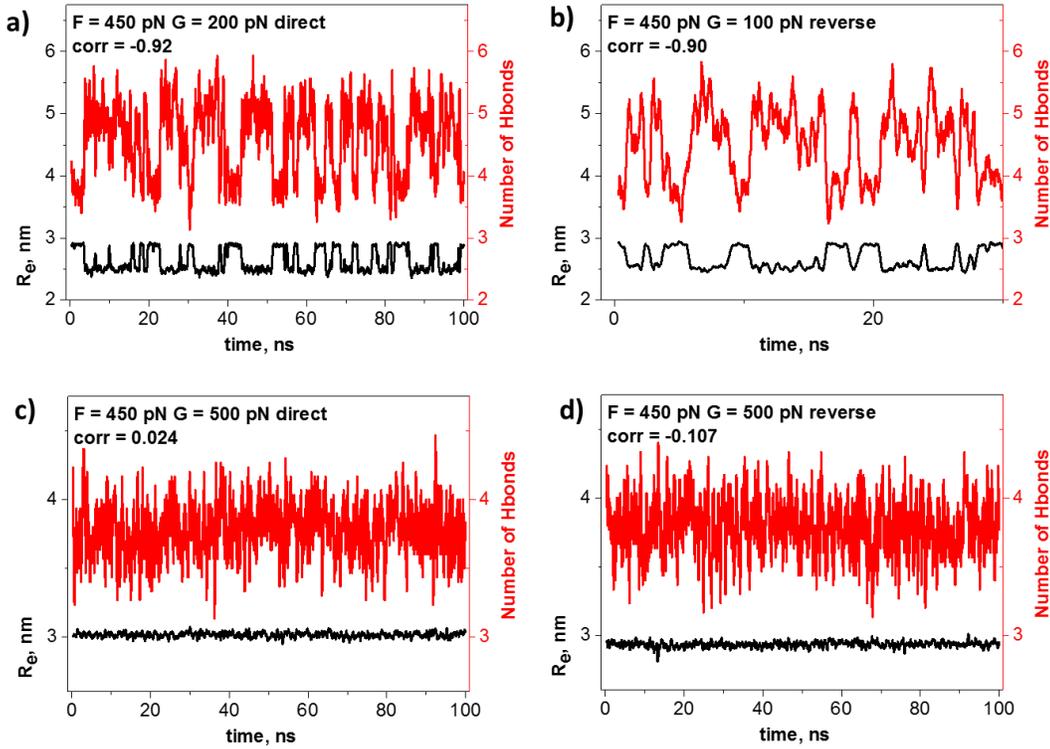

**Figure 5.** Time series of the end-to-end distance $R_e(t)$ and the number of hydrogen bonds located around of the oligomer bending area (HBBA). a) Low-frequency vibrations of $R_e(t)$ near the $G_1$ critical point (black curve; left axis) correlate with the low-frequency vibrations of HBBA (red curve; right axis); the correlation coefficient is equal to $-0.92$. b) Low-frequency vibrations of $R_e(t)$ near the $G_2$ critical point (black curve; left axis) correlate with the low-frequency vibrations of HBBA (red curve; right axis); the correlation coefficient is equal to $-0.92$. c) and d) There were no appreciable correlations between $R_e(t)$ and HBBA far from the critical lateral loads: for $G = 500$ pN, the correlation coefficients were equal to 0.024 and $-0.107$, respectively.

For both critical values $G_1 = 100$ pN and $G_1 = 200$ pN, we observed pronounced vibrations of the oligomer and the number of HBBA, which are synchronized in antiphase with high correlation coefficients of $-0.90$ and $-0.92$, respectively (Figures 5a,b). Figures 5c,d illustrate that the vibrations are absent far from the critical points and that the time series $R_e(t)$ does not correlate with the time series HBBA. According to these data, the durations of the vibration pulses were also roughly 10 ns. As a result, the activation barrier for transitions between the straightened and bended shapes ranged from 10–15 $kT$. Therefore, the thermal-activated switching of hydrogen bonds between the olio-30s-NIPMAm and environmental water appeared to be a common mechanism of low-frequency, mechanic-like vibration, regardless of whether the vibration referred to emergence of the soft dynamical mode near the threshold longitudinal loads or to the robust dynamical modes under the catastrophes near the critical lateral loads. In fact, this effect appears to be a size effect; dynamical bistability may be activated by thermal fluctuations in such small, non-linear systems.

In conclusion, our computer simulations have revealed that the oligomers of thermosensitive polymers a few nanometers in size can respond to power loads exactly like an Euler arch. In other words, short oligomers may act as catastrophe machines. We wish to emphasize that the oligo-30s-NIPMAm we established in this letter is not a unique sample. We found a number of oligomers and block-oligomeric compositions of thermosensitive polymers with properties of a nanoscale Euler arch. In general, we suggest that oligomers (or oligomeric compositions) consisting of several Kuhn segments with lengths not significantly less than 1 nanometer can function like classical catastrophe machines if such oligomers (or oligomeric compositions) possess the characteristics of bistable nonlinear systems.

**Supporting Information:**

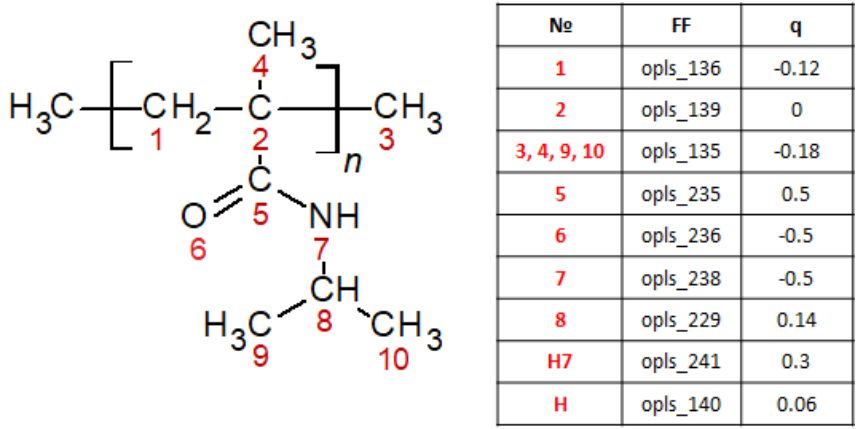

Figure S1. OPLS-AA force field parameters of molecular dynamic simulations of N-isopropylmethylacrylamid.

ACNOWLEDGMENTS

We thank Sergei Nechaev and Alexander Gorsky for helpful discussions. This work was conducted in the framework of a project supported by the Molecular Machine Corporation.

REFERENCES

1. Zeeman, E. C. A catastrophe machine. In *Towards a Theoretical Biology.* Waddington C. H., Ed., Edinburg University Press, Edinburg, 1968, 4, pp. 276-282.
2. Arnol'd, V. I. *Catastrophe Theory.* Springer-Verlag: Berlin, Heidelberg, 1984, IX; p. 79.
3. Poston, T.; Stewart I. *Catastrophe Theory and Its Applications.* Courier Corporation, 1996, p. 491.
4. Cleland, A. N. *Foundation of Nanomechanics. From Solid-State Theory to Device Applications.* Springer-Verlag: Berlin Heidelberg, 2003, XII; p. 436.
5. *Nanotribology and Nanomechanics I. Measurment Technique and Nanomechanics.* Bhushan B., Ed., Springer-Verlag: Berlin Heidelberg, 2011, XVII; p. 623.
6. Zhang, L.; Marcos, V.; Leigh, D. *Molecular machines with bio-inspired mechanisms.* Proc. Nat. Acad. Sci. USA, 2018, 115(38), pp. 9397-9404.
7. Van Dijk, L.; Tilby, M. J.; Szpera, R.; Smith, O. A.; Bunce, H. A. P.; Fletcher, S. P. *Molecular machines for catalysis.* Nature Reviews Chemistry, 2018, 2, pp. 0117-0120
8. Credi, A.; Venturi, M. Molecular machines operated by light. Central European Journal of Chemistry. 2008 6(3), pp. 325-339.
9. Bulsara, A. R.; Schieve, W. C.; Gragg, R. F. *Phase transitions induced by white noise in*
10. *bistable optical systems*. Phys. Lett. A. 1978, 68, pp.. 294-296.
11. Horsthemke, W.; Lefever, R. *Noise-Induced Transitions. Theory and Applications in Physics, Chemistry, and Biology*. Springer-Verlag: Berlin, Heidelberg, 1984, XVI; p. 322.
12. Halperin, A., Kröger, M.; Winnik, F. Poly(N-isopropylacrylamide) *Phase Diagrams: Fifty Years of Research,* Angewante Chemie (International ed. in English), 2015, 54, pp. 15342-15367.

13. Hoogenboom, R. *Tunable Thermoresponsive Polymers by Molecular Design*. In *Complex Macromolecular Architectures: Synthesis, Characterization, and Self-Assembly*. Hadjichristidis, N., Hirao, A., Tezuka Y., and Du Prez F., Eds., Wiley & Sons: New York, 2011, Chapter 22, pp. 685-715.
14. Berendsen, H.J.V.; van der Spoel, D.; van Drunen, D. *GROMACS: A message-passing parallel molecular-dynamics implementation*. Comp. Phys. Comm., 1995, 91, pp. 43-56.
15. Lindahl, E.; Hess, B.; van der Spoel D. *GROMACS 3.0: a package for molecular simulation and trajectory analysis*. J. Mol. Model., 2001, 7, pp. 306-317.
16. Van Der Spoel, D.; Lindahl, E.; Hess, B.; Groenhof, G.; Mark, A. E.,; Berendsen, H. J. C. *GROMACS: Fast, flexible, and free*. Journal of Computational Chemistry, 2005, 26(16), pp. 1701-1718.
17. Walter, J.; Sehrt, J.; Vrabec, J.; Hasse, H. *Molecular Dynamics and Experimental Study of Conformation Change of Poly(N-isopropylacrylamide) Hydrogels in Mixtures of Water and Methanol*. J. Phys. Chem. B, 2012, 116, pp.. 5251-5259.
18. De Oliveira, T.E.; Mukherji, D.; Kremer, K.; Netz, P. *Effects of stereochemistry and copolymerization on the LCST of PNIPAm*. J Chem Phys., 2017, 146, pp. 034904-034912.
19. Alaghemandi, M.; Spohr, E. Molecular dynamics investigation of the thermoresponsive polymer poly(N-isopropylacrylamide). Macromol Theory Simul., 2012, 21, pp. 106-112